\documentclass[a4paper,11pt]{article}
\usepackage{pos}

\title{sPHENIX measurements of heavy flavor production in $p$+$p$ collisions}

\author*[a]{Xudong Yu}
\onbehalf{On behalf of the sPHENIX Collaboration}

\affiliation[a]{School of Physics,Peking University,\\
  209 Chengfu Road, Beijing, People's Republic of China}

\emailAdd{yuxd@stu.pku.edu.cn}

\abstract{sPHENIX is the first new collider detector experiment dedicated to heavy-ion physics since the LHC began collecting data. Successfully commissioned in 2023–2024, one of its standout features is a streaming-capable tracking system that enables the collection of large, unbiased p+p datasets—previously unattainable at the Relativistic Heavy Ion Collider (RHIC). Leveraging this capability, sPHENIX recorded over 100 billion unbiased $p$+$p$ collisions at 200 GeV during Run 24. This unprecedented dataset unlocks a high-precision open heavy flavor physics program with extended low-$p_T$ reach, spanning both charm and beauty sectors. These proceedings present the progress in the analysis of open heavy flavor in the $p$+$p$ dataset. From one hour of data and early-stage calibrations, we see observations of $D^0$ mesons and evidence of $\Lambda_c^+$ in $p$+$p$ collisions for the first time at RHIC. These resonances will allow for novel physics measurements to be performed for the first time at RHIC.}

\FullConference{The European Physical Society Conference on High Energy Physics (EPS-HEP2025)\\
7-11 July 2025\\
Marseille, France\\}

\begin{document}
\maketitle

\section{Introduction}
The sPHENIX experiment is a state-of-art collider detector built at the Relativistic Heavy Ion Collider (RHIC) at Brookhaven National Laboratory~\cite{PHENIX:2015siv}. Its primary physics goal is to explore the properties of the Quark Gluon Plasma (QGP) using hard probes, such as jets and heavy flavor particles, in the low $p_T$ region where RHIC production is more abundant, thereby complementing studies performed at the Large Hadron Collider (LHC). Due to their large masses, heavy quarks are produced during the early stages of hard scattering in collisions, and thus experience the entire evolution of the QGP medium as well as the hadronization process. The sPHENIX detector is specifically designed to measure hadrons containing heavy flavor charm and bottom quarks. A key objective of the experiment is to determine the $\Lambda_c^+/D^0$ yield ratio in $p$+$p$ collisions, which provides new insights into the hadronization mechanism.

The construction of sPHENIX was completed in 2023, with commissioning carried out using Au+Au beams in 2023, followed by $p$+$p$ collisions in 2024. After the completion of commissioning phase, sPHENIX successfully collected a large sample of unbiased $p$+$p$ collision data, taking advantage of its unique streaming readout capabilities. In these proceedings, we first provide a brief overview of the sPHENIX detector, then detail the performance of the tracking reconstruction, and finally present the current status of ongoing heavy flavor analysis at sPHENIX.

\section{sPHENIX detector and tracking reconstruction}

The sPHENIX detector consists of a high-precision tracking system, large-acceptance electromagnetic and hadronic calorimeters within the central barrel, complemented by forward and endcap detectors including the Minimum Bias Detector (MBD), the Event Plane Detector (sEPD), and Zero-Degree Calorimeter (ZDC). The central barrel provides coverage of $|\eta| < 1.1$ with full azimuth, while the MBD, sEPD, and ZDCs are located outside this region. Four dedicated tracking detectors are used for tracking reconstruction and heavy flavor analysis. The Monolithic Active Pixel Sensors (MAPS)-based Vertex detector (MVTX) is a precision vertex detector composed of three layers of MAPS staves located a few centimeters of the beam pipe. The MVTX provides a hit position resolution down to 5 microns, with a vertex resolution down to approximately 10 microns. This precise vertexing capability is essential for identifying heavy flavor particles with short decay lengths. The Intermediate silicon Tracker (INTT) is a two-layer silicon strip detector. Its key capability is a timing resolution sufficient to resolve individual 106.5 ns RHIC bunch crossings in $p$+$p$ collisions, enabling the correct association of tracks and mitigating pile-up effects. The Time Projection Chamber (TPC) is a gaseous detector with 48 layers spanning radii from 20 cm to 78 cm and equipped with GEM-based continuous readout. It provides the momentum resolution required for the sPHENIX tracking physics program. Although the TPC is not designed for particle identification (PID), its ionizing measurement $dE/dx$ offers excellent separation power at low momentum, which is particularly useful for suppressing the combinatorial background in heavy flavor analysis. The TPC Outer Tracker (TPOT) is an eight-tile Micromegas detector and provides additional hit position measurement outside the TPC. The primary purpose of the TPOT is to enable correction of the space charge distortions in the TPC. Together, these four detectors provide essential information for the four-dimensional reconstruction of tracks, allowing the determination of both the spatial and temporal structure of physics events. Tracks are reconstructed through a multi-stage process: clustering hits in individual detectors, seeding in the silicon and TPC subsystems separately, matching track seeds in space and time, and performing a final Kalman Filter fit using the Acts software package~\cite{Osborn:2021zlr,Ai:2021ghi}.

\begin{figure}
    \centering
    \includegraphics[width=0.5\linewidth]{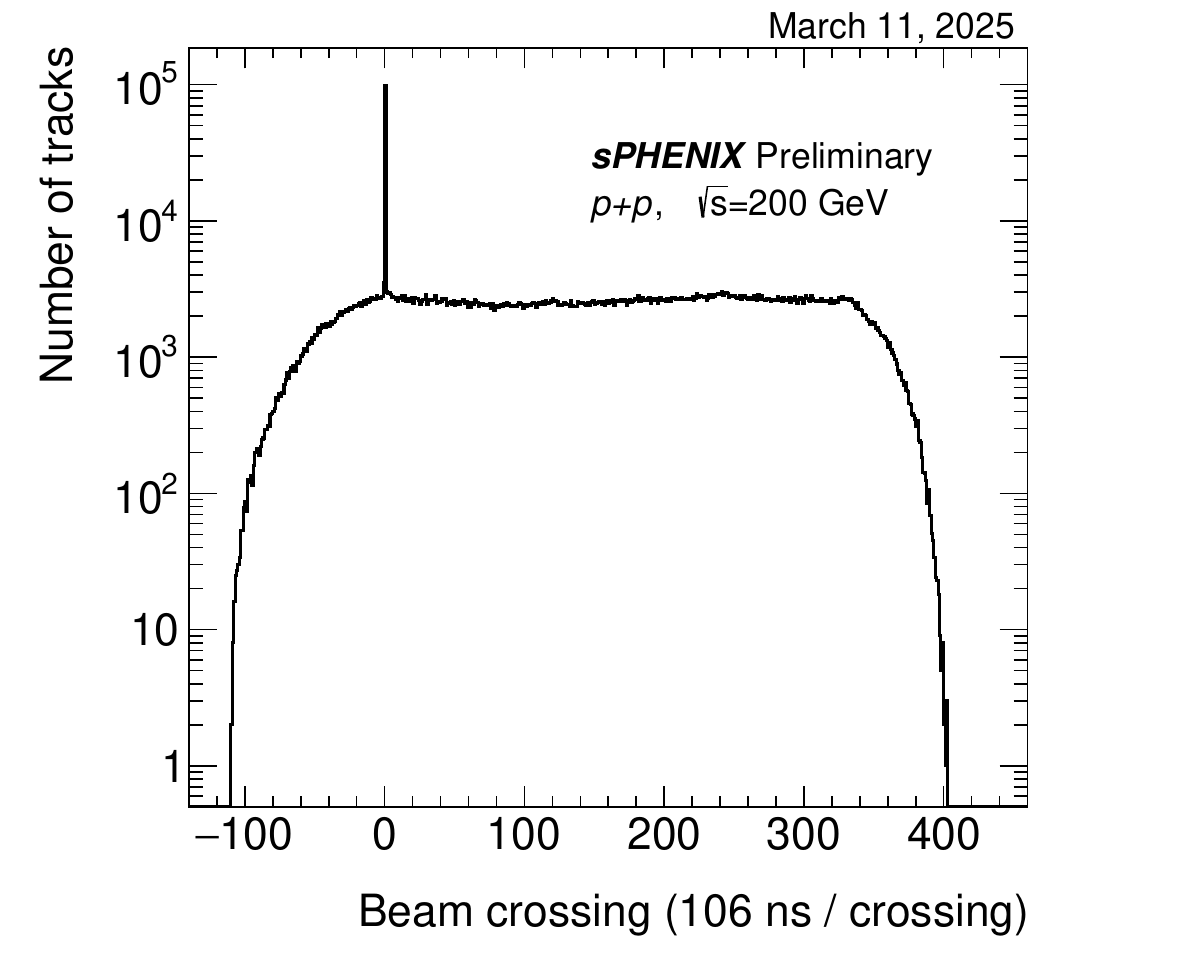}
    \caption{The number of reconstructed tracks in $\sqrt{s}=200$ GeV $p$+$p$ collisions as a function of the relative beam crossing number, where crossing 0 corresponds to the hardware trigger timing.}
    \label{fig:crossing}
\end{figure}

The sPHENIX tracking detectors operate in a unique extended streaming readout mode, in which approximately 20\% of the $p$+$p$ collisions delivered by RHIC are recorded. The motivation for adopting the streaming readout mode arises from the open heavy flavor physics program, where low $p_T$ open heavy flavor hadrons suffer from poor hardware trigger efficiencies. By the end of the Run 24 $p$+$p$ data taking period, the tracking detectors had recorded unbiased $p$+$p$ collisions at a rate of $\mathcal{O}(200)$ kHz, which is about 20-50 times higher than the rate achieved with a pure minimum-bias trigger. Figure~\ref{fig:crossing} shows the number of reconstructed tracks as a function of the beam crossing number relative to the hardware trigger, where a crossing number of 0 is coincident with the hardware trigger, and all other crossings represent tracks reconstructed from the streamed data. An approximately constant tracking efficiency is observed across all crossings, demonstrating that the streaming readout data acquisition and reconstruction performed successful. Over the entire Run 24 period, sPHENIX collected 100 billion streaming events, equivalent to an integrated luminosity of 2.9 pb$^{-1}$.

\section{Heavy flavor reconstruction}
\begin{figure}
    \centering
    \includegraphics[width=0.4\linewidth]{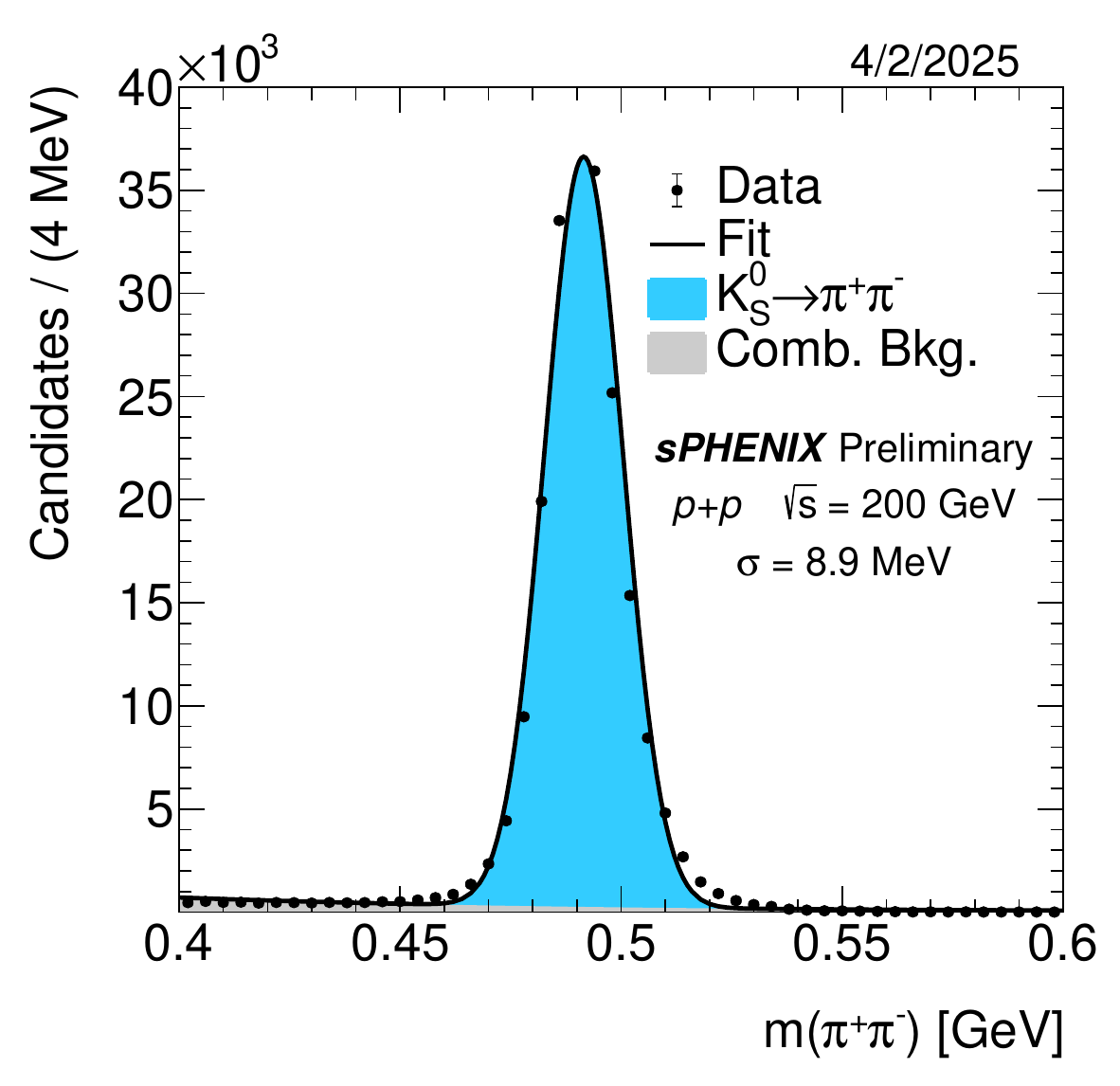}
    \includegraphics[width=0.4\linewidth]{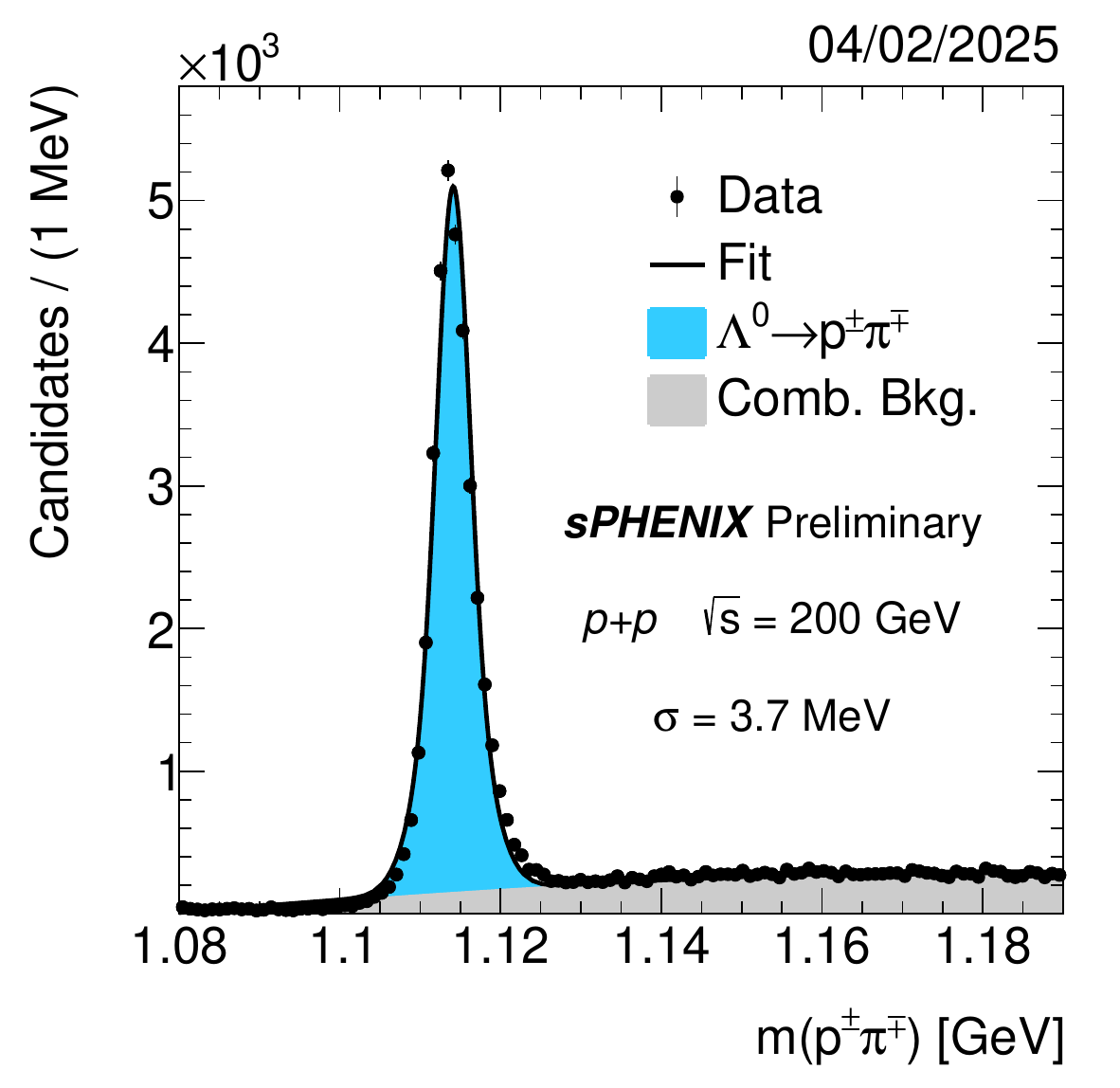}
    \includegraphics[width=0.4\linewidth]{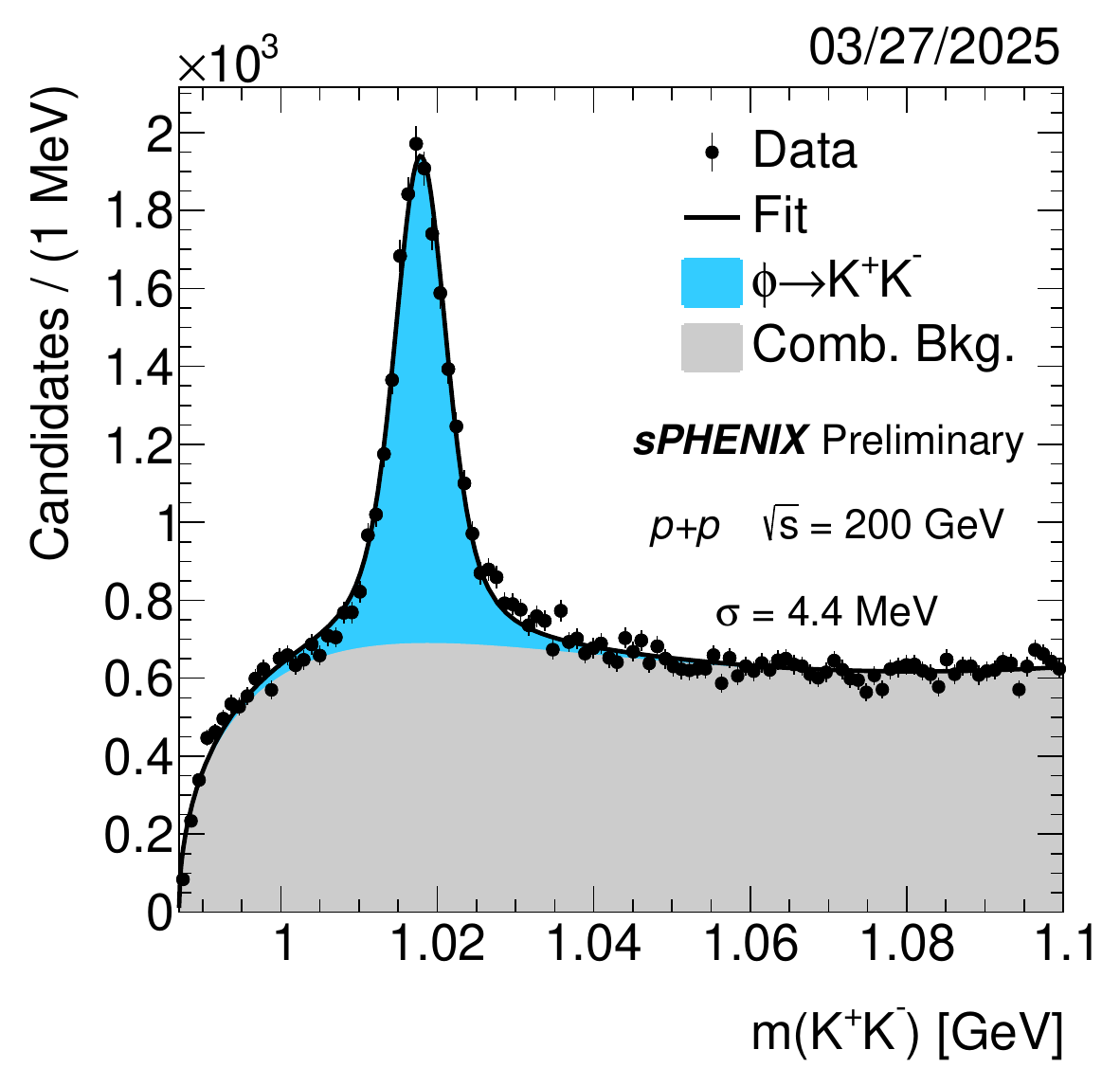}
    \includegraphics[width=0.4\linewidth]{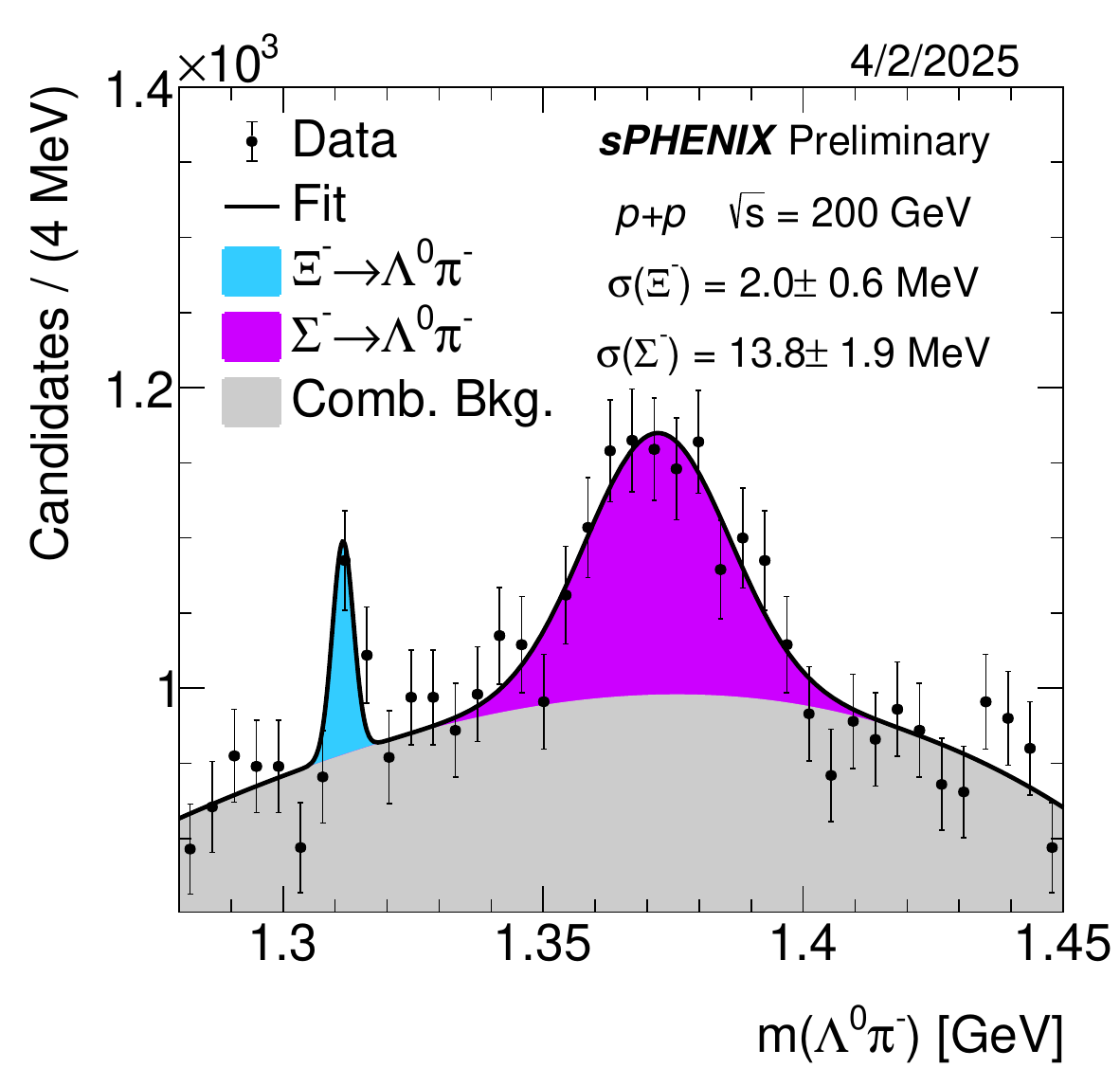}
    \caption{The invariant mass of reconstructed $\pi^+\pi^-$, $p\pi^-$, $K^+K^-$, and $\Lambda\pi^-$ combinations. The combinatorial background is shown in gray, and the signals of $K_S^0$, $\Lambda$, $\phi$, $\Xi^-$ and $\Sigma(1385)^-$ are shown in blue or purple.}
    \label{fig:lf}
\end{figure}

In the months following the completion of Run 24 data taking, the analysis efforts have focused on a small subset of tracking data, corresponding to approximately one hour of integrated luminosity. Extensive work is ongoing to understand and calibrate the first full collision data collected by the sPHENIX tracking detectors, with particular emphasis on the alignment of the silicon detector using field-off data and on the correction of TPC space charge distortions based on the lamination fitting. After applying initial calibrations, many well-known light flavor resonances can be clearly observed. Examples include the $K_S^0$ reconstructed via $\pi^+\pi^-$ decay channel, the $\Lambda$ via $p\pi^-$, the $\phi$ via $K^+K^-$, and even more complex three-body final states such as $\Xi^-/\Sigma(1385)^-\to\Lambda\pi^-$ with $\Lambda\to p\pi^-$, as shown in Figure~\ref{fig:lf}. The reconstruction of these resonances is performed using the KFParticle package, which has been adapted to the sPHENIX software stack~\cite{kfparticlewiki}. This serves as an important validation exercise for the overall tracking and vertexing performance prior to tackling more challenging heavy flavor measurements.

\begin{figure}
    \centering
    \includegraphics[width=0.4\linewidth]{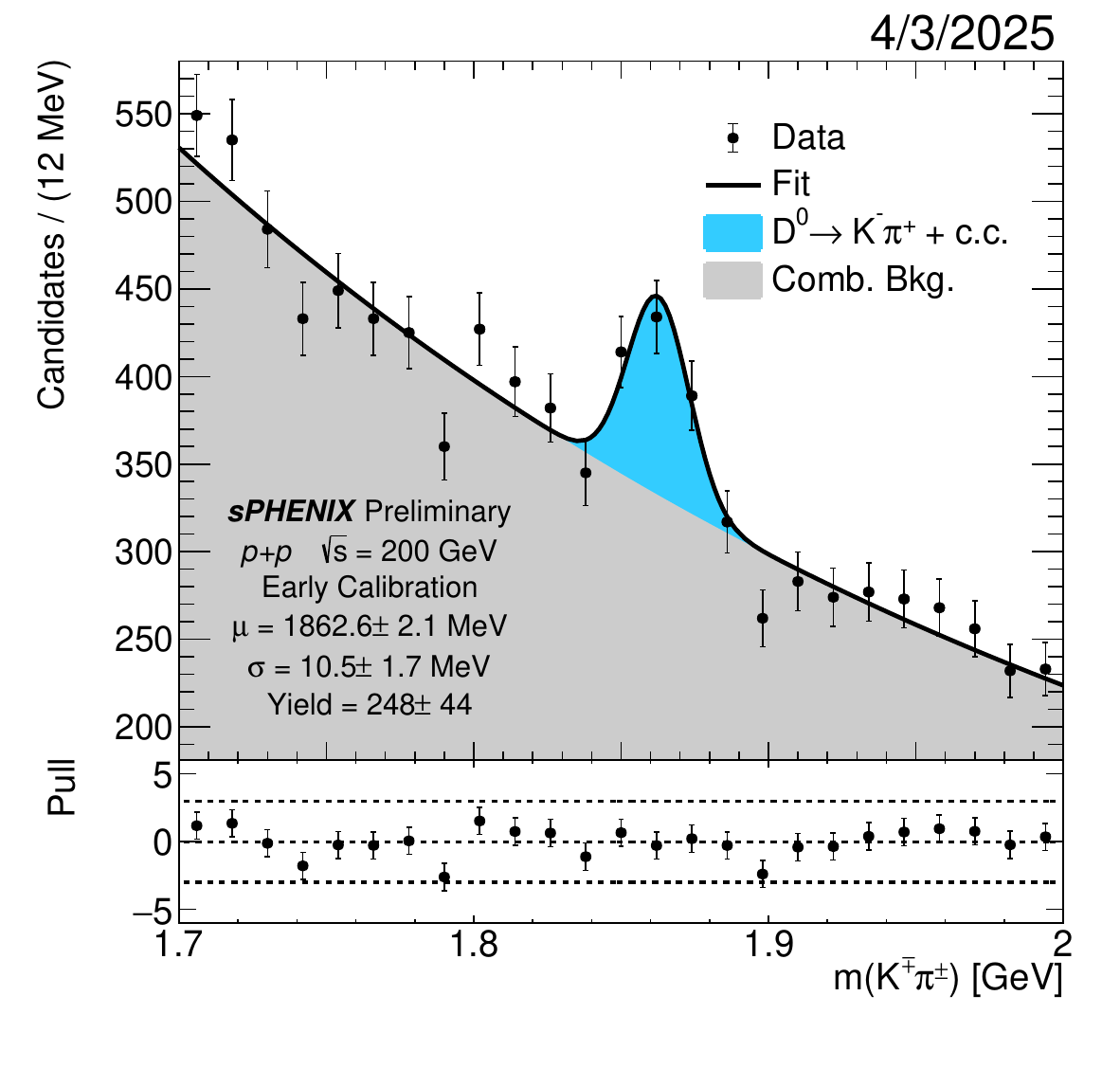}
    \includegraphics[width=0.4\linewidth]{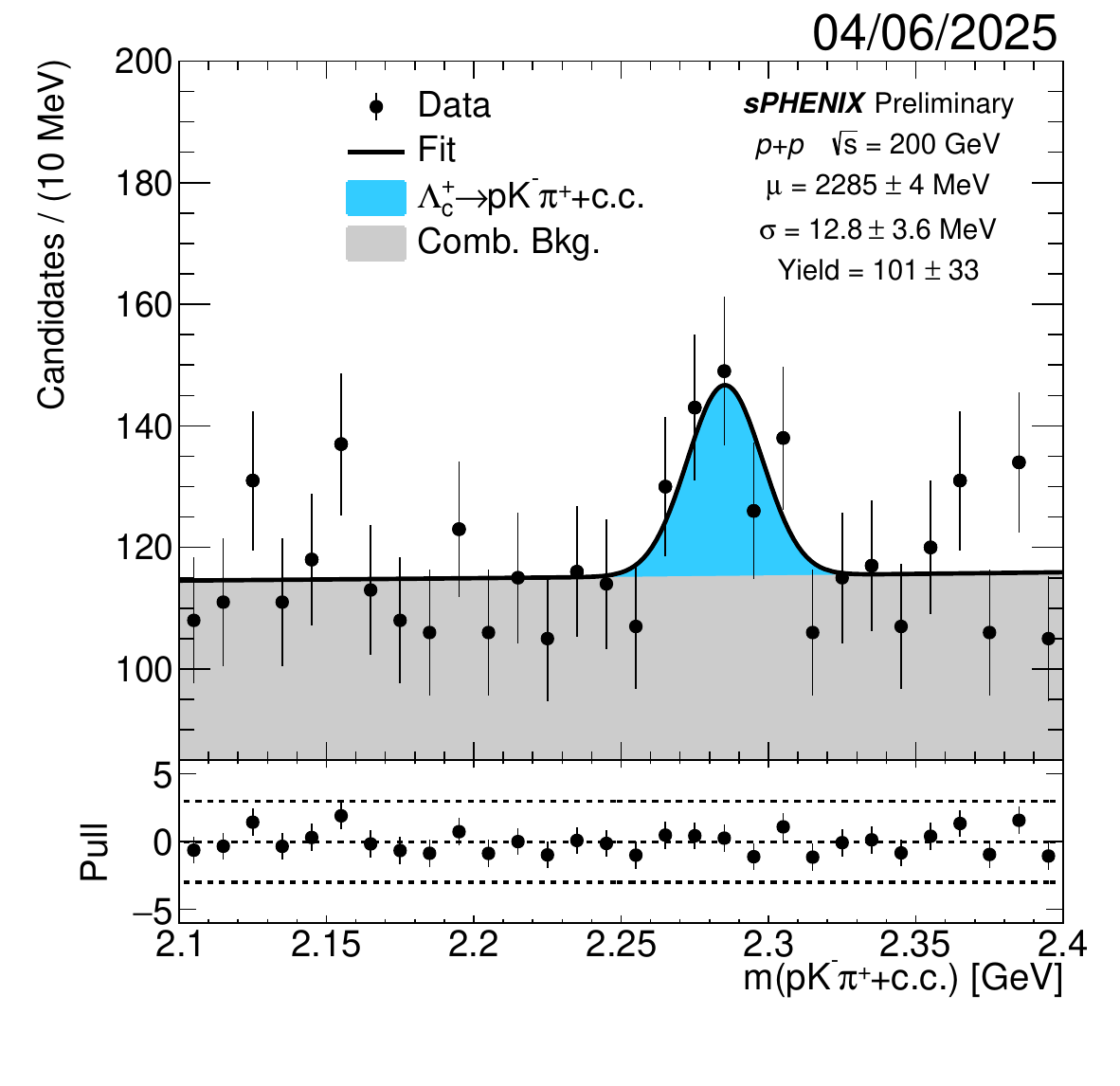}
    \caption{(Left) The invariant mass of the reconstructed $K^-\pi^+$ pairs. (Right) The invariant mass of $pK^-\pi^+$ combinations. The combinatorial background is shown in gray, and the signals of $D^0$ meson and $\Lambda_c^+$ baryon are shown in blue.}
    \label{fig:hf}
\end{figure}

Compared with light flavor resonances, heavy flavor states possess larger masses, lower production yields, and are subject to significantly higher combinatorial backgrounds, making their reconstruction considerably more demanding. Figure~\ref{fig:hf} (left) presents the first invariant mass peak measurement of the $D^0$ meson, reconstructed through $D^0\to K^-\pi^+$ channel, observed with a significance of approximately 5.6$\sigma$ at sPHENIX. Figure~\ref{fig:lf} (right) shows the first observation in $p$+$p$ collisions at RHIC of an invariant mass peak of the $\Lambda_c^+$ baryon, reconstructed via the $pK^-\pi^+$ channel. The reconstructed secondary vertex and low-momentum $dE/dx$ information from the TPC are utilized to suppress the huge amounts of combinatorial background, although the track pointing resolution and momentum resolution remain limited by the current knowledge of silicon detector alignment and TPC distortion corrections. The ongoing calibration efforts aim to further improve the precision of both the reconstructed vertex and momentum measurements, which are the essential prerequisites for an accurate determination of the $\Lambda_c^+/D^0$ yield ratio.

\section{Summary}
The sPHENIX experiment has collected 100 billion unbiased $p$+$p$ events at $\sqrt{s}=200$ GeV druing Run 24. The tracking detector, featuring a streaming readout system and advanced reconstruction capability, has successfully demonstrated the measurement of a wide range of resonances~\cite{sphenixpublic}, including the first measurement of $D^0$ meson from sPHENIX and the first measurement of $\Lambda_c^+$ in $p$+$p$ at RHIC. sPHENIX is actively taking a large Au+Au dataset in 2025, aiming for an integrated luminosity of 7.0 nb$^{-1}$. This dataset will enable a rich program of heavy flavor physics, providing precision measurements to further elucidate the properties of the QGP.

\end{document}